\documentclass[aps,showpacs,floatfix,superscriptaddress,twocolumn]{revtex4-1}
\usepackage{graphicx}
\usepackage{amssymb}
\usepackage{graphics, color}
\usepackage{amsmath}
\usepackage{overpic}
\usepackage{float}
\usepackage{hyperref}
\usepackage{enumitem}

\usepackage{amsmath}

\graphicspath{{pics/}}

\definecolor{Gray}{gray}{0.4}

\begin{document}
\title{Comment on \href{http://arxiv.org/abs/1301.3829}{1301.3829} and Reply to `Comment on ``Can disorder really enhance superconductivity?"' (\href{http://arxiv.org/abs/1501.05148}{1501.05148})
 by I.M. Suslov}
\author{James Mayoh}

\author{Antonio M. Garc\'\i a-Garc\'\i a}
\affiliation{TCM Group, Cavendish Laboratory, University of Cambridge, JJ Thomson Avenue, Cambridge, CB3 0HE, UK}

\begin{abstract}
In \cite{suslov,suslov1} I. M. Suslov proposes a theoretical description for the interplay between disorder and superconductivity that, among other things, claims to predict situations where superconductivity is enhanced by disorder. In this comment we show that Suslov's results do not make any sound predictions relating to this problem. Suslov also suggests that the percolation approach recently employed by the authors to compute the global critical temperature of inhomogeneous superconductors induced by weak multifractality \cite{us} is not satisfactory. We refute Suslov's opinion using simple ideas from the Bardeen-Cooper-Schiaffer (BCS) theory of superconductivity. More importantly we stress that our theory makes quantitative analytical predictions for observables such as the global critical temperature and the spatial distribution of the order parameter. Our work can therefore be easily confirmed or disproved through either numerical or experimental study.  
\end{abstract}
\date{\today}

\maketitle

Suslov has recently claimed \cite{suslov1} to have developed a theory, for the interplay of superconductivity and disorder, that is capable of answering questions like: Can disorder enhance superconductivity? He also criticised \cite{suslov} the percolation approach proposed by the authors\cite{us} to address this problem in the case of metallic superconductors with inhomogeneity induced by weak multifractality. In the following we refute this criticism and show that Suslov's theory is based on an ill defined critical temperature and the unjustified use of the mean-field approximation. Unsurprisingly, Suslov's main prediction of a large enhancement of the critical temperature driven by disorder has never, and very likely will never, been found numerically or experimentally.

Suslov studies the critical temperature of a superconductor $T_c$ after adding a small
concentration of point-like impurities in two limits: 
\begin{enumerate}
  \item {\bf Strong coupling}: corresponding to an impurity that act as a deep potential well of the size the lattice spacing, $a$, and is effectively isolated from the rest of the system. This region is completely unrelated to our recent investigation \cite{us}.
  \item {\bf Weak coupling}: corresponding to a weak random potential where the superconductor is not strongly affected by the impurity.
\end{enumerate}
Let us first discuss the case of strong coupling. Suslov assumes that these impurities induce discrete levels around the Fermi energy and then uses mean-field techniques to compute the critical temperature. His main result is that $T_c \sim E_0 \lambda$ where $E_0$ is of the order of Fermi energy and $\lambda$ is the electron-phonon coupling constant. However, it is well established that a mean-field approximation is completely unjustified in this ultra-small limit where the effective mean-level spacing $\delta$, controlled by the lattice spacing $a$, is much larger than the bulk gap $\Delta_0$.

 In this region,  rightly termed  {\it catastrophic} by Suslov, the spectroscopic gap must be computed using either the Richardson equation \cite{richardson} or perturbative techniques \cite{larkin} starting from the normal state.
More importantly, the critical temperature $T_c$ is ill defined in this fluctuation dominated regime \cite{scalapino} where the spectroscopic gap at the Fermi energy is always non-zero at finite temperature \cite{scalapino,ribeiro}. Even for $\Delta_0 \sim \delta$ \cite{scalapino,ribeiro} there is no substantial decay of the spectroscopic gap at the mean-field critical temperature. This is not surprising since for sizes comparable to the lattice spacing, superconductivity is only a weak perturbation \cite{larkin}. 
Suslov's prediction of $T_c \sim 5000$K, for values corresponding to typical metallic superconductors, is therefore completely, though not surprisingly, unrealistic.

Suslov goes on to argue, reasonably, that since this critical temperature only occurs in a few, well separated regions, of typical size given by the lattice spacing $a$, it cannot be the true global critical temperature of the sample. More specifically this $T_c$ is not the one at which global phase-coherence, and zero resistance, occurs. Suslov claims that it is possible to tune the number of resonant impurities, controlled by  $T_c/E_0$, by simply increasing the density of impurities, so that,
``{\it eventually their Josephson interaction is strong enough for stabilization of the mean-field solution at
practically the same $T_c$ value}".
We first note this heuristic picture of emergent granularity for establishing phase-coherence in a disordered superconductor was proposed a long time ago \cite{ma,trivedi} and is still actively pursued \cite{trivedi1}. However, the way Suslov implements it is open to criticism. First, there is broad evidence that this appealing picture is correct starting from the insulator side and gradually reducing disorder so that phase coherence is eventually achieved.
  Surprisingly Suslov applies it in the opposite direction since, according to him, phase coherence occurs by gradually {\it increasing} disorder. It is unclear to us how  any genuine realisation of disorder can have this property. Disorder makes the system more localized and therefore always increases the magnitude of phase fluctuations in the system.  We stress that as disorder decreases from the deep insulator limit superconducting spots start to appear but are related to relatively flat regions of the disorder not to any {\it resonant scattering} \cite{trivedi}. 
  
Even assuming Suslov's picture and its prediction of $T_c$ to be accurate, which we have shown above is not the case, 
in a weakly coupled superconductor $\lambda \ll 1$ so $T_c/E_0$ is always small and the global critical temperature must necessarily be zero. Moreover we have serious reservations as to whether it is possible to tune this number in any realistic disordered material.
 In summary, Suslov's statement that, after including phase fluctuations, the global critical temperature is still not very different from $T_c \sim 5000$K is absolutely unrealistic.

Towards the end of \cite{suslov} Suslov argues, and we agree, that Coulomb interactions will eventually reduce the global critical temperature to a value much smaller than $T_c \sim E_0 \lambda \ll 5000$K. Unfortunately he does not give any quantitative or semi-quantitative estimate of the suppression of the critical temperature due to this effect. The suppression of $T_c$ will strongly depend on the typical size of the superconducting drops and the coupling strength between neighbouring grains. There is recent experimental evidence that even for almost isolated superconducting nano-grains of Sn and Pb \cite{sangita} of size $\sim 7$nm there is no Coulomb blockade or even substantial reduction of the superconducting gap. It is therefore unlikely that Coulomb interactions can reduce the critical temperature by the several orders of magnitude Suslov's theory implies, at least for metallic superconductors. This is another indication that Suslov's critical temperature, $T_c \sim E_0 \lambda$, is ill defined and not directly related to a typical energy of any realistic superconductor.
 
 In summary the enhancement of superconductivity predicted by Suslov in the {\it catastrophic regime} is unphysical. Not surprisingly it has never been observed and, from our point of view, it will never be observed numerically or experimentally. We must also emphasise that this model is completely unrelated to the weakly coupled inhomogeneous superconductor induced by weak multifractality that we have recently investigated \cite{us}.\\

We now turn to the weakly coupled disordered region of Ref.\cite{suslov,suslov1}. According to Suslov the critical temperature is given by,   
 \begin{equation}
 \frac{\delta T_c}{T_{c0}} = \frac{1}{\lambda L^d}\,
\int d^dr \, \frac{ \nu_0  \nu_1(r) +  \nu_1(r)^2}{\nu_0^2}
\,, \label{weak}
 \end{equation}
where   $\nu_1({r})$ is the deviation of the local 
density of states at the Fermi energy, $\nu_F(r)$, from its unperturbed value $\nu_0$,
$\lambda=g\nu_0$ is the dimensionless coupling constant,
$T_{c0}$ is the transition temperature in the absence of disorder and $L^d$ is ``{\it a volume for one impurity}". The length, $L$, is poorly defined as it is unclear whether it is related to the system size. Similarly  $\nu_1({r})$ is not properly defined as a function 
of a small parameter. On physical grounds we speculate that its strength is controlled by the dimensionless conductance, $g_0$, so that the small parameter in the expansion is $1/g_0\lambda$. Irrespective of this, the simple dependence on $\lambda$ indicates that this is a first order purely perturbative result that describes the leading correction to the Anderson theorem. 
Suslov uses this expression as the starting point for his criticism of our recent paper \cite{us}, in which we compute analytically the global critical temperature of a disordered superconductor induced by weak-multifractlity. 

Before we address, and refute, his criticism we summarise our main results. In Ref.\cite{us} we compute
the superconducting order parameter analytically, including its energy dependence and statistical
distribution in space. The spatial distribution of the order parameter is found to be always log-normal. By applying percolation techniques we determine the global critical temperature neglecting
phase fluctuations. We find that it is enhanced with respect to the clean limit only for very weakly
coupled superconductors. This enhancement seems robust as it persists even after the percolation threshold is increased in order to crudely estimate the effect of phase-fluctuations. It is important to note that in a disordered superconductor there are different critical temperatures: the global critical temperature corresponding to the maximum at which a supercurrent can flow, the local critical temperature at which the order parameter at a given point vanishes and the critical temperature related to the vanishing of the gap at the Fermi energy. For instance the first is the natural outcome of a transport measurement and the second is the natural outcome of a scanning tunnelling microscope experiment \cite{sangita}. One of the conclusions of our recent paper \cite{us} is that a sharp division between these distinct, but related, critical temperature is a fundamental requirement if one is to make substantial progress on this problem.

The main criticism of Ref. \cite{suslov} is directed at the percolation approach we use to compute the global critical temperature. According to Suslov, since we are working in the weakly disordered regime of Eq. (\ref{weak}), the local critical temperatures $T_c({\bf r})$, defined as the temperature at which the amplitude of the order parameter at $r$ vanishes, depends only weakly on $r$. Consequently the value of $T_c$ resulting from percolation is similar to that obtained by averaging over $T_c({\bf r})$ and therefore a percolation approach is not necessary. This is not correct. Below we argue that Suslov's opinion is based on some misunderstanding of basic features of our model and of Bardeen-Cooper-Schrieffer (BCS) theory as well.

It is well known \cite{mirlin} that multifractality, a key ingredient in our results, induces strong correlations \cite{mirlin,kravtsov} among eigenstates with energies within the Debye window, $\epsilon_D$. These enhance the spatial inhomogeneities of the order parameter even in the limit of weak-multifractality; corresponding to a large dimensionless conductance. Moreover, in  BCS theory, the order parameter depends exponentially on the coupling constant, $\lambda$. As a consequence, small corrections to the coupling constant, due to disorder or other perturbations, induce an exponential effect on the order parameter provided that the calculation is carried out self-consistently, typically involving a re-summation of perturbative corrections. This is why even weak perturbations can induce strong modifications of the gap and critical temperature. We must also stress that there is substantial experimental \cite{sangita}, numerical \cite{shanenko,shanenko1} and analytical \cite{altshuler,us0} evidence that supports this statement. 

The $1/\lambda$ dependence of Eq. (\ref{weak}) reveals a simple first order perturbative result that does not take into account the non-perturbative effect of the correlations between multifractal eigenstates.  By contrast the combination of multifractality and a fully selfconsistent analytical calculation leads to a broad log-normal distribution
for the spatial distribution of the superconducting gap, the associated local critical temperature, as well as to a much more involved  dependence of the global critical temperature on the coupling constant $\lambda$. It is evident, see Fig $4$ of Ref. \cite{us}, that such a broad distribution leads to important variations of $T_c({\bf r})$ at different points in space. Moreover we are interested only in weakly coupled superconductors in the limit of weak perturbations where a mean-field approach is still acceptable and the necessary truncation of the Hilbert space, by expressing the matrix elements as a function of the one-body eigenstates, is still a good approximation \cite{shanenko,shanenko1}. This fully justifies the computation of an upper bound for the global critical temperature by a percolation analysis of $T_c({\bf r})$. Indeed our results, see Fig. $4$ of Ref. \cite{us}, indicate the global critical temperature is vastly different, and more importantly lower, than any averaged $T_c({\bf r})$ equivalent to $T_c$ in Suslov's paper, even before any reduction due to Coulomb interactions.

We also believe part of Suslov's criticism of Ref. \cite{us} is simply based on a poor understanding of our percolation approach. According to Suslov our percolation analysis is intended to describe fluctuations. This is not the case. In a first approximation our percolation analysis neglects {\it completely} phase fluctuations. It only gives the maximum temperature for which a finite {\it amplitude} of the superconducting gap permeates the sample. We then go on to provide a rough estimation of the effect of phase fluctuations by increasing the percolation threshold so that, even if superconducting regions permeate the sample, there is no long-range order. This is the correct approach when starting from a purely metallic state which becomes increasingly inhomogeneous with increasing disorder. Indeed there is recent experimental evidence \cite{pratap} that the spatial distribution of the order parameter in a metallic quasi two-dimensional disordered superconductor is similar to the one we derived analytically. \\

Finally we comment on other statements in Suslov's papers that are not correct.
Suslov claims that our results are only valid in small and strictly two dimensional systems. However it is well established in the theory of disordered systems that weak multifractality also occurs in quasi two dimensional systems (thin films) and two dimensional systems with spin-orbit interactions. All of these systems are of direct experimental relevance. 

Suslov also states, referring to our calculation of the critical temperature,  that "{\it there is no need to use the approximate equation (6)} [equation (6) in \cite{suslov}] {\it when the accurate result} [restated here as equation \ref{weak}~] {\it is available}". We emphasise that precisely the opposite is true. Suslov's formalism is strictly perturbative with no sharp distinction between the different critical temperatures that define a disordered superconductor. Self-consistency is not reinforced and consequently the dependence on $\lambda$ is trivial. Eq. \ref{weak} could be a good starting point to re-derive our results. However this will require much more effort than Ref. \cite{suslov} suggests. 

Suslov claims towards the end of \cite{suslov} that we do not discuss phase-fluctuations and Coulomb interactions. We must point out that section VII from Ref.\cite{us} {\it does} already discuss this in some detail from both a qualitative and semi-quantitative viewpoint. 
Suslov also states that we claim $E_0$ is always much larger than $\epsilon_D$. We are interested in weakly coupled  metallic superconductors so in our case this limit is always applicable. However, we make no such claim regarding other materials. 

An important part of Ref. \cite{suslov} is actually devoted to criticising other recent publications on superconductivity with multifractal eigenstates \cite{kravtsov,mirlin} that are not directly related to our work. For example, the critical temperature at which the gap at the Fermi energy vanishes is, in these publications, $T_c \sim E_0 \lambda^{1/\gamma}$ with $\gamma$ related to the multifractal dimensions. This is an expression that, we are at pains to stress, is not recovered in our formalism in the limit of weak-multifractility. \\

To summarise, Suslov's claim to have explored conclusively the possibility for enhancement of superconductivity due to disorder are clearly unfounded. His theory is based on ill defined concepts, the results are rarely testable and lead, in general, to unphysical predictions. Similarly, the criticism of the percolation formalism employed in our paper \cite{us} is not warranted. Percolation is a powerful tool for an quantitative study of the problem of inhomogeneous superconductivity induced by weak-multifractality. 
 
 \vspace{1cm}
 
 \bibliography{library}


\end{document}